\def\year{2017}\relax
\documentclass[letterpaper]{article}
\usepackage{aaai17}
\usepackage{times}
\usepackage{color}
\usepackage{xcolor}
\usepackage{booktabs}
\usepackage{helvet}
\usepackage{hyperref}
\usepackage{courier}
\usepackage{tabu}
\usepackage{booktabs}
\usepackage{amssymb}
\usepackage{graphicx}
\usepackage{hyperref}
\usepackage{mathtools}
\usepackage{textcomp}
\usepackage{subfig}
\usepackage{amssymb}

\long\def\ignore#1{}

\newcommand{\rev}[1]{{\color{red}{#1}}}

\begin{document}
\title{Creative Community Demystified: A Statistical Overview of Behance}
\author{Nam Wook Kim\\
Harvard University\\
}

\maketitle
\begin{abstract}
Online communities are changing the ways that creative professionals such as artists and designers share ideas, receive feedback, and find inspiration. While they became increasingly popular, there have been few studies so far. In this paper, we investigate Behance, an online community site for creatives to maintain relationships with others and showcase their works from various fields such as graphic design, illustration, photography, and fashion. We take a quantitative approach to study three research questions about the site. What attract followers and appreciation of artworks on Behance? what patterns of activity exist around topics? And, lastly, does color play a role in attracting appreciation? In summary, being male suggests more followers and appreciations, most users focus on a few topics, and grayscale colors mean fewer appreciations. This work serves as a preliminary overview of a creative community that later studies can build on.

\end{abstract}

\section{Introduction}

Online communities are flourishing in recent years, enabling creative practitioners such as artists and designers to share ideas, receive feedback, and find inspiration across the globe, which is a critical part of their professional practice~\cite{tan2015destuckification,salah2012deviantart,perkel2011making}. Well-known creative community sites include Dribble\footnote{\url{www.dribbble.com/}}, Behance\footnote{\url{www.behance.net}}, and DeviantArt\footnote{\url{www.deviantart.com}}. In contrast to other online social networks targeted for general audiences such as Facebook, Twitter, and Pinterest, these sites are specifically designed for people who work in creative fields, specifically for visual arts. They have greatly influenced creatives on publicizing their works and connecting with like-minded people, which in turn helped democratizing the arts.
 
Such specialized communities provide unique opportunities for understanding the new practice of producing and sharing creative artworks and how social connections among creative professionals are structured. They are rather different from social curation sites such as Pinterest, where users collect, organize, and share collections of random items, in that contents shared in the creative communities are made by users themselves. These users are mostly concerned about promoting their original works, finding other inspiring works, and finding other talented people. 

In this paper, we investigated Behance, an online community site where users showcase and discover creative works from various fields such as graphic design, illustration, photography, and fashion. Each user has a unique personal page showing collections of artworks that they \textit{authored} or \textit{appreciated}. Behance users can follow other users as a way of subscribing to the works of the others. Behance is one of the largest creative community sites and has recently drawn some research interest~\cite{halstead2015finding,rudolph2016joint}; however, there is so far no in-depth user behavior analysis available.

This work aims to provide a comprehensive statistical overview of Behance. To guide our research, we defined three specific questions below.

\begin{description}
 \item[R1-Activity] What drives user activity? For example, what attracts appreciations or followers? Do women or men attract more attention?
 \item[R2-Topic] What is the topical structure of Behance? For instance, what topics are most popular and how they are related? Do people tend to appreciate works created by people with similar interests?
 \item[R3-Color] Does it matter what colors are used for artworks? That is, do certain colors attract more appreciations of the artworks?
\end{description}
\vspace*{-2mm}

We used quantitative methods to study these questions, conducting a statistical analysis of data sampled from the Behance network using publicly available API. Our research questions and analysis methods are significantly inspired by existing research on Pinterest that has similar features and network structures~\cite{chang2014specialization,gilbert2013need,bakhshi2015red}. 

In short, we found that female users attract fewer followers and appreciations, most users specialize in a few similar topics, and grayscale colors mean fewer appreciations. Our results offer a glimpse of overall user activity on Behance and have implications for creative communities in general.

\section{Background}
\subsection{Online Creative Communities}
Creative professions include a variety of fields including not only art and design, but also writing, crafting, theater, marketing, and even scientific research~\cite{florida2012rise}. The scope of this work lies in online communities specifically designed for visual arts ranging from traditional art forms such as drawing, painting, and architecture to applied arts such as graphic design, fashion design, and photography. 

Many online creative communities emerged over the past decade with the widespread use of web technologies. They significantly influenced the creative practice of producing and sharing artworks by promoting a participatory structure surpassing the proprietary nature of traditional art practice~\cite{perkel2011making}. They helped artists expand the audience for their work, find inspirational people, distribute artwork to continue to be reproduced \& redistributed~\cite{perkel2011making}.

While the impact of such online communities is not negligible, there have been few scholarly studies. Moreover, most existing studies are qualitative research~\cite{salah2010online,perkel2011making,scolere2016pinning,jones2015collective}. \citeauthor{perkel2011making} conducted an extensive ethnographic study of devianART, a massive online community site built around sharing digital art, and investigated user behavior in relation to specific features of the site. \citeauthor{scolere2016pinning} studied the implications of curatorial labor on Pinterest for creative professionals. Some studies take data-driven approaches to study devianART, involving network, image, and text analysis methods~\cite{AkdagSalah2013,salah2012deviantart}, but the depth of their data analysis is mostly shallow.

\subsection{User Behavior Analysis on Social Networks}

Over the last few years, online social networks have become central places for people to maintain social relationships and share similar interests. Much research has been devoted to analyzing user behavior on such social network sites including Twitter~\cite{kwak2010twitter}, Tumblr~\cite{chang2014tumblr}, Instagram~\cite{hu2014we}, Pinterest~\cite{gilbert2013need}, and many others. 

So far, no data analysis on the same level has been conducted for social network services designed for creative communities. Existing work on devianART is mostly qualitative, while studies on Dribble~\cite{deka2015ranking} and Behance are focused on ranking~\cite{halstead2015finding} and recommendation~\cite{rudolph2016joint}.

Behance has features and network structures similar to Pinterest, a popular social curation site where people collect, organize, and share image-based contents. For this reason, existing studies on Pinterest shed some light on what users would behave on Behance, and thus we attempt to leverage on the studies in this work. 

\citeauthor{gilbert2013need} did a quantitative study providing a statistical overview of Pinterest. Later, \citeauthor{chang2014specialization} looked at the extent to which users specialize in particular topics, and homophily among users, while \citeauthor{ottoni2013ladies} did a deeper analysis of the role of gender. \citeauthor{han2014collecting} similarly analyzed curation patterns in relation to topics and gender. On the other hand, \citeauthor{bakhshi2015red} investigated image contents, mainly about how color is related to the diffusion of pins (i.e., repins). Our work borrows some of the research questions and analysis methods from these works and applies to Behance.

\section{Platform Description}

\begin{figure}[tb]
  \includegraphics[width=\linewidth]{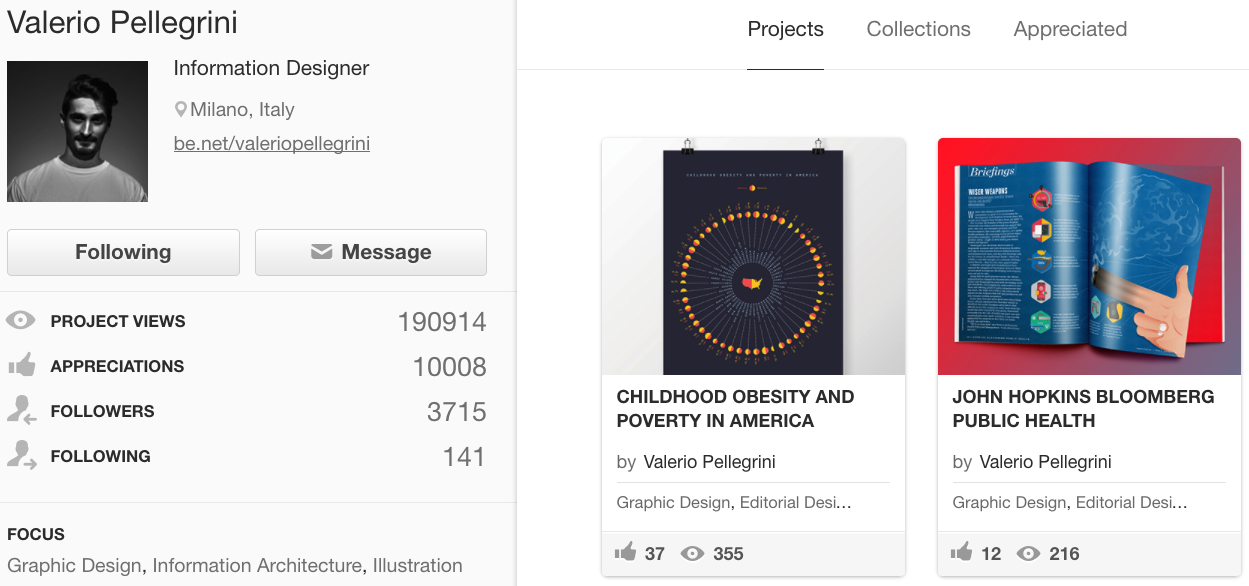}
  \caption{A profile page of a famous infographic designer. On the left, basic information is shown, including total counts of appreciations and followers and specialization topics. On the right, a collection of projects authored by the designer is listed.  }
  \label{fig:profile}
  
\end{figure}

Behance is an online community site for people to showcase and discover creative works as well as to connect to other people with similar interests. The goal of the site in its own words is to democratize opportunities for talented creative professionals\footnote{\url{https://www.behance.net/about}}, serving as a social portfolio site. Contents shared on the site are original works authored by users themselves who want to promote their works.  

Figure~\ref{fig:profile} shows a portion of a user profile page. 
A user can create a collection of projects which appear in the profile page along with other basic information such as a residency location and stats about the user's activity; \text{project} is the term used to refer to a creative work on Behance. A user can specify up to three topics for each project. The user's focus is automatically generated based on the topics most used in the projects; for instance, the user in the Figure~\ref{fig:profile} specializes in graphic design, information architecture, and illustration. 

A user can also engage in social interactions by viewing, appreciating, or commenting on other people's projects. The list of projects appreciated by the user also appears on the profile page (implicit curation). The user can also organize collections of projects that they find interesting (explicit curation), similar to Boards on Pinterest. However, this is not actively used as the goal of the site is to promote users' own works; e.g., we observed a median collection count of zero per user in our dataset.

Similar to other online social networks, a user can follow other users as a way of subscribing to the works of the others which appear in the user's private timeline. The relationship between users is not symmetric similar to Pinterest, suggesting that the Behance network is also interest-driven. Users can send direct messages to each other, enabling private interaction. 

There is public timeline as well, allowing for serendipitous discovery. In the timeline, users can browse projects by topics, social metrics (e.g., most appreciated), and even geographic locations. They can also browse people and teams using the same criteria.

\section{Data}
In this section, we describe the data we obtained from Behance, and how we obtained and processed it to prepare for our analysis. To answer our research questions, we collected data about users and projects. For users, we were interested in data that indicated their activities. For projects, we gathered data allowing us to trace who created and appreciated them in what topics and what color information are used for project images. Table~\ref{tbl:data}. shows an overview of the data used in our analysis.

\begin{table}[tb]
\centering
\caption{A list of variables for user and project data used in our analysis.}
\label{tbl:data}
\begin{tabu} to 1.0\linewidth{X[0.5,l] X[1.5,l]}
\toprule
\textit{User}                  &                  \\ \midrule
Gender & Inferred from a user's name.            \\
Focus          & A user's specialized topics \\
Country        & A user's main place of residence               \\
Followers       & \# of users follow this user                   \\
Following & \# of users this user follows \\
Appreciations  & \# of users liked this user's projects \\ 
Comments  & \# of comments on this user's projects \\ 
Counts  & \# of designs created by this user\\ 
Views  & \# of users viewed this user's projects \\ \midrule
\textit{Project}                  &                  \\\midrule
Appreciation & \# of users liked this project\\
Topics & Topics to which this project belongs\\
H,S,V & Avg.pixels in the HSV color space \\
W3C Colors & \% of pixels close to 16 W3C Colors\\
Colorfulness1 & Defined by \citeauthor{hasler2003measuring}\\
Colorfulness2 & Defined by Yendrikhovskij et al.\\
\bottomrule
\end{tabu}
\end{table}

\subsection{Gathering the data}
Since the whole Behance network is extremely large to deal with, our goal was to obtain a random sample of users and projects using the official API provided by Behance\footnote{\url{https://www.behance.net/dev}}. We used the sampling approach employed by~\citeauthor{chang2014specialization}. It is based on the random walk sampling proposed by~\citeauthor{leskovec2006sampling} with a simple modification that a seed set consists of active users drawn from a public timeline. 

We sampled Behance data on Nov 2016, performing the random walk starting from a public timeline page listing most recent projects. We randomly picked a user from a seed set of users whose projects appeared in the public timeline; thus the seed users are assumed to be active users. We randomly chose the next user from the followers and followees of the user and then repeated the process from the chosen user until a desired number of users are collected. With the probability of 0.15, the process randomly jumped back to the starting user in the seed set. To avoid being stuck in sink nodes, we did a random jump after every 1000 steps of walking the network by again randomly choosing another starting user in the seed set. 

Once we collected a total of 50,000 users from the random walk, we extracted 881,208 projects and 2,608,052 appreciations from the users. We limited the number of appreciations to at most 80 per user since the number can be tremendously large and thus intractable.

\subsection{Preparing the data for analysis}

\begin{figure}[tb]
  \includegraphics[width=\linewidth]{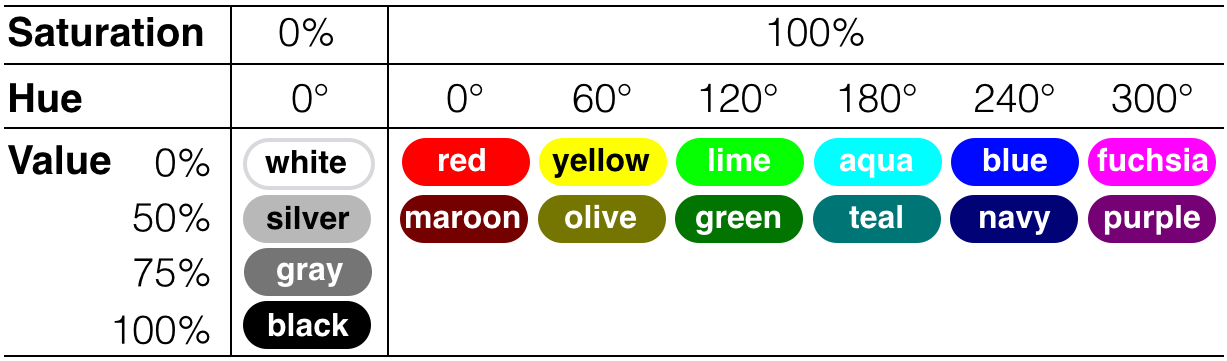}
  \caption{16 W3C Colors}
  \label{fig:w3s-colors}
\end{figure}

Once we collected the data described above, we further did additional work to prepare it for analysis. 

In order to analyze gender roles, we used a commercial service\footnote{\url{www.gender-api.com}} to infer a user's gender based on the user's first name and country of residence if available; Behance does not require users to specify their gender. To ensure the accuracy of our analysis, we used gender predictions only when the prediction precision is higher than 90\%; when the only first name is available, we used a 95\% precision threshold. 

To generate color information for each project, we additionally crawled project images from Behance using URLs available in the project data we sampled. We only extracted thumbnail images whose resolution is 200x158 in order to speed up the color extraction process. 

We used the same color metrics defined in~\citeauthor{reinecke2013predicting}, including 1) the percentage of 16 colors defined by the W3C (Figure~\ref{fig:w3s-colors}), 2) the average pixel value for hue, saturation, and value, 3) colorfulness1 measuring the color difference against gray using the weighted sum of the trigonometric length of the standard deviation in ab space and the
distance of the center of gravity in ab space to the neutral axis ~\cite{hasler2003measuring}, and 4) colorfulness2 as the sum of the average saturation value and its standard deviation where the saturation is computed as chroma divided by lightness in the CIELab color space~\cite{yendrikhovskij1998optimizing}. The colorfulness metrics were originally designed for natural images, and both showed a high correlation with human ratings~\cite{reinecke2013predicting}.

Among the total of 50,000 users, 26,653 users are male (53\%), 11,124 users are female (22\%), and 12,223 users are unknown gender (24\%). In our analysis, we only used \textit{37,777 users} (75\%) that have gender information, resulting in \textit{668,581 unique projects} created by the users, and \textit{1,534,032 appreciations} made to the projects by the same users. For color analysis, the number of projects was trimmed down to 542,345 due to failures of downloading images and extracting color information.

\section{Method}
\subsection{Activity Analysis}
We used negative binomial regression to model user activity on Behance as a function of the predictive variables listed in Table~\ref{tbl:data} (\textit{User}). We used follower and appreciation counts as our dependent variables as they are strong social signals for user activity and popularity ~\cite{deka2015ranking,kwak2010twitter}. We used negative binomial regression instead of Poisson regression since the variance of each dependent variable is larger than the mean (Follower: $\mu$=1.30K, $\sigma$=6.70K, appreciation: $\mu$=2.47K,  $\sigma$=8.88K); negative binomial regression is better suited for over-dispersed distributions of count dependent variable~\cite{cameron2013regression}. To incorporate categorical variables into our model, we constructed binary country and topic variables such as \textit{from united states} and \textit{from graphic design}, similar to~\citeauthor{gilbert2013need}. We evaluated each model by comparing it to an intercept-only (null) model and examined the reduction in deviance.

\subsection{Topic Analysis}
To analyze the topical structure of Behance, we mainly drew on methods used by~\citeauthor{chang2014specialization}.

First, we represented all projects that users created and appreciated in our dataset using a total of 68 topics provided by Behance. We defined the topic vector of a project as a binary vector as below.

\[
    \vec{d}_i=<d_{i,1}, d_{i,2},...,d_{i,68}>
\]

where, $d_{i,j}$ is zero or one indicating the appearance of $j$th topic on project $d_i$. The topic vector of our whole dataset is then represented as a normalized count vector of the topics of all projects in the dataset:

\[
    \hat{t}_{\boldsymbol{C}}=\frac{\vec{t}_{\boldsymbol{C}}}{\left\lVert\vec{t}_{\boldsymbol{C}}\right\rVert_1},\textrm{where}\quad   \vec{t}_{C}=\sum_{\vec{d_i}\in \boldsymbol{C}}{\vec{d}_i}
\]
where $j$th value in the vector indicates the proportion of $j$th topic for all projects. Similarly, we also represented each user by aggregating the topic vectors of all the projects created by the user:

\[
    \hat{u}_{\boldsymbol{C}}=\frac{\vec{u}_{\boldsymbol{C_u}}}{\left\lVert\vec{u}_{\boldsymbol{C_u}}\right\rVert_1}, \quad \textrm{where} \quad  \vec{u}_{\boldsymbol{C_u}}=\sum_{\vec{d_i}\in \boldsymbol{C_u}}{\vec{d}_i}
\]
 and $\boldsymbol{C_u}$ is a set of projects that the user $u$ created. We similarly defined $\hat{u}_{\boldsymbol{A_u}}$, where $\boldsymbol{A_u}$ consists of projects that the user $u$ appreciated.
 
 To investigate how topics are related, for each topic $i$, we constructed $T_i$ a binary vector of the size of the total number of projects we collected. A non-zero value in the $j$-th position in the vector denotes that project $j$ has topic $i$. We computed Jaccard distance between two topics, where a 0 indicates perfect similarity and a 1 means total dissimilarity; we used the Jaccard distance as $T_i$ contains many zero entries.
 
 \[
    sim(T_i,T_j) = \frac{| T_i \cap T_j |}{| T_i \cup T_j |}
 \]
 
 This enables us to compute pairwise distances for all 68 topics, measuring the frequency of co-occurrence of two topics. We also performed a hierarchical clustering of the pairwise distances using Ward's criterion minimizing the with-in cluster variance.
 
 To analyze how diverse topics of projects are created per user, we calculated the entropy of a user's topic vector as follows:
 
 \[
    Entropy(\hat{u}) = -\sum_{j=1}^{68}u_j \log_e{u_j}
 \]
 where $u_j$ is the $j$th entry in a user vector $\hat{u}_{\boldsymbol{P}}$, denoting the fraction of projects in $j$th topic.
 
 Finally, we also wanted to analyze homophily of interests. We were specifically interested in the extent to which users appreciate projects in the same topics that they specialize in. As our measure of homophily, we used the cosine similarity of a user's two topic vectors constructed from two sets of projects that the user created and appreciated respectively.
 
 \[
    homophily(\hat{u})=cosine(\hat{u}_{\boldsymbol{C_u}}, \hat{u}_{\boldsymbol{A_u}}) = \frac{\hat{u}_{\boldsymbol{C_u}}\cdot \hat{u}_{\boldsymbol{A_u}}}{|\hat{u}_{\boldsymbol{C_u}}|\cdot |\hat{u}_{\boldsymbol{A_u}}|}  
 \]
 While we do not directly compute homophily using the relationship between users such as using followers and followees, our measure serves the same purpose since the appreciated projects were created by other users.
 
\subsection{Color Analysis}
 Contents shared on Behance are driven by the visual images of projects. In this work, we examined colors used in the images, one of the most prominent image features. We were interested in the effect of color on the appreciation of projects; that is, are certain colors more conducive to draw attention from users?
 
 We followed~\citeauthor{reinecke2013predicting} to extract various color metrics including 16 W3C colors, average values of hue, saturation, and value, and two colorfulness measures. Similar to \citeauthor{bakhshi2015red}, we used negative binomial regression using the number of project appreciations as the dependent variable. Our predictor variables include not only the color variables but also a control variable (mean follower count of owners of a project). We looked at the reduction in deviance from the full model to the control-only model to test the significance of the color variables on explaining the number of project appreciations. 
 
 While \citeauthor{bakhshi2015red} picked dominant hues (binary variables), our W3C color variables are numeric values and thus can handle multiple dominant colors in an image. The W3C color variables can be considered as representative colors sampled from a 3-dimensional HSV color cube; their coverage of hue ranges from 0\textdegree to 360\textdegree with 60\textdegree interval (Figure~\ref{fig:w3s-colors}). 

\section{Results}
\subsection{Descriptive Statistics}

\begin{table}[tb]
\centering
\caption{Basic statistics of the variables for a total of 37,777 users. All quantitative variables have zero as their minimum. Only top ranking countries and specializations are included for categorical variables. }
\label{tbl:stats}
\begin{tabu} to 1.0\linewidth{X[l] X[0.2,r] X[0.3,r] X[0.3,r] X[0.3,r]}
\toprule
\textit{Numeric Variable} & Q1 & Median & Mean & Q3 \\ \midrule
followers & 15 & 86 & 1.3k & 560 \\
following & 38 & 134 & 455.5 & 408 \\
appreciations & 20 & 156 & 2.5k & 1.3k \\
views & 279 & 2.3k & 32.1k & 14.8k\\
counts & 4 & 12 & 17.95 & 22\\
comments & 1 & 11 & 150.8 & 92\\ \midrule
 \textit{Categorial Variable} &  &  &  &  \\ \midrule
Gender & \multicolumn{2}{l}{26.7k male} & \multicolumn{2}{l}{11.1k female} \\
Country & \multicolumn{2}{l}{5.4k USA} & \multicolumn{2}{l}{3.2k Brazil} \\
 & \multicolumn{2}{l}{2.0k UK} & \multicolumn{2}{l}{1.6 Russia} \\
 & \multicolumn{2}{l}{1.4k Italy} & \multicolumn{2}{l}{1.3k India} \\
Specialized Topics & \multicolumn{4}{l}{16.6k graphic design} \\
 & \multicolumn{4}{l}{12.1k illustration} \\
 & \multicolumn{4}{l}{6.9k art direction} \\ \bottomrule
\end{tabu}
\end{table}

Table~\ref{tbl:stats} reports basic statistics of the user variables in our dataset. As similar to other social network data, the distributions of variables are skewed, containing a significant number of zeros. For example, the differences between medians and means are considerably large for follower and appreciation counts. All the variables have zero minimums and large portions of them are zeros (e.g., follower: 4.95 \%, appreciation: 13.28\%). 

The number of male users is more than two times higher than the number of female users. This is noteworthy compared to Pinterest, where female users dominate the overall population (almost more than 90\%)~\cite{chang2014specialization}. Many users specialize in graphic design (43.9\%), illustration(32.1\%), and art direction (18.3\%); the focus topics were automatically calculated by Behance based on the topics of the users' projects. \citeauthor{chang2014specialization} reported that male users are more interested in "Design" than female users on Pinterest, which is one possible explanation for the gender distribution in our dataset. 

Roughly 14.4\% of users are from the United states\ignore{ particularly from California and New York(\rev{xx})}. The top three countries (USA, Brazil, and UK) remain the same for both female and male users. Since we only consider users of whom we were accurately able to predict gender, many users from Asian countries were excluded (e.g., China).

\subsection{R1-What attracts appreciations?}

\begin{table}[tb]
\centering
\caption{The result of a negative binomial regression with a number of \textit{appreciations} as the dependent variable. The table only lists a subset of predictors in the order of relative importance based on standardized coefficients. However, the $\beta$ coefficients here remain on their original scales.}
\label{tbl:nb1}
\begin{tabular}{@{}lllll@{}}
\toprule
Predictor & Estimate & std.err & z & p \\ \midrule
intercept & 4.47 & $2.42$e$^{-2}$ & 184.9 & \textless $2$e$^{-16}$ \\
comments & $2.61$e$^{-3}$ & $2.35$e$^{-5}$ & 111.2 & \textless $2$e$^{-16}$ \\
views & $1.03$e$^{-5}$ & $1.07$e$^{-7}$ & 96.10 & \textless $2$e$^{-16}$ \\
counts & $1.81$e$^{-2}$ & $2.45$e$^{-4}$ & 73.86 & \textless $2$e$^{-16}$ \\
has illustration & $4.52$e$^{-1}$ & $2.14$e$^{-2}$ & 21.07 & \textless $2$e$^{-16}$ \\
has art direction & $5.38$e$^{-1}$ & $2.14$e$^{-2}$ & 21.07 & \textless $2$e$^{-16}$ \\
following & $8.33$e$^{-5}$ & $4.17$e$^{-6}$ & 20.00 & \textless $2$e$^{-16}$ \\
has typography & $6.42$e$^{-1}$ & $3.37$e$^{-2}$ & 19.07 & \textless $2$e$^{-16}$ \\
followers & $2.47$e$^{-5}$ & $1.67$e$^{-6}$ & 14.82 & \textless $2$e$^{-16}$ \\
has digital art & $4.77$e$^{-1}$ & $2.69$e$^{-2}$ & 17.71 & \textless $2$e$^{-16}$ \\ \midrule
Summary & null.dev & res.dev & $\chi^{2}$ & p \\\midrule
 & 84.5K & 47.3K & 37.2K & \textless $2$e$^{-16}$ \\ \bottomrule
\end{tabular}
\end{table}

\begin{figure}[b]
  \includegraphics[width=\linewidth]{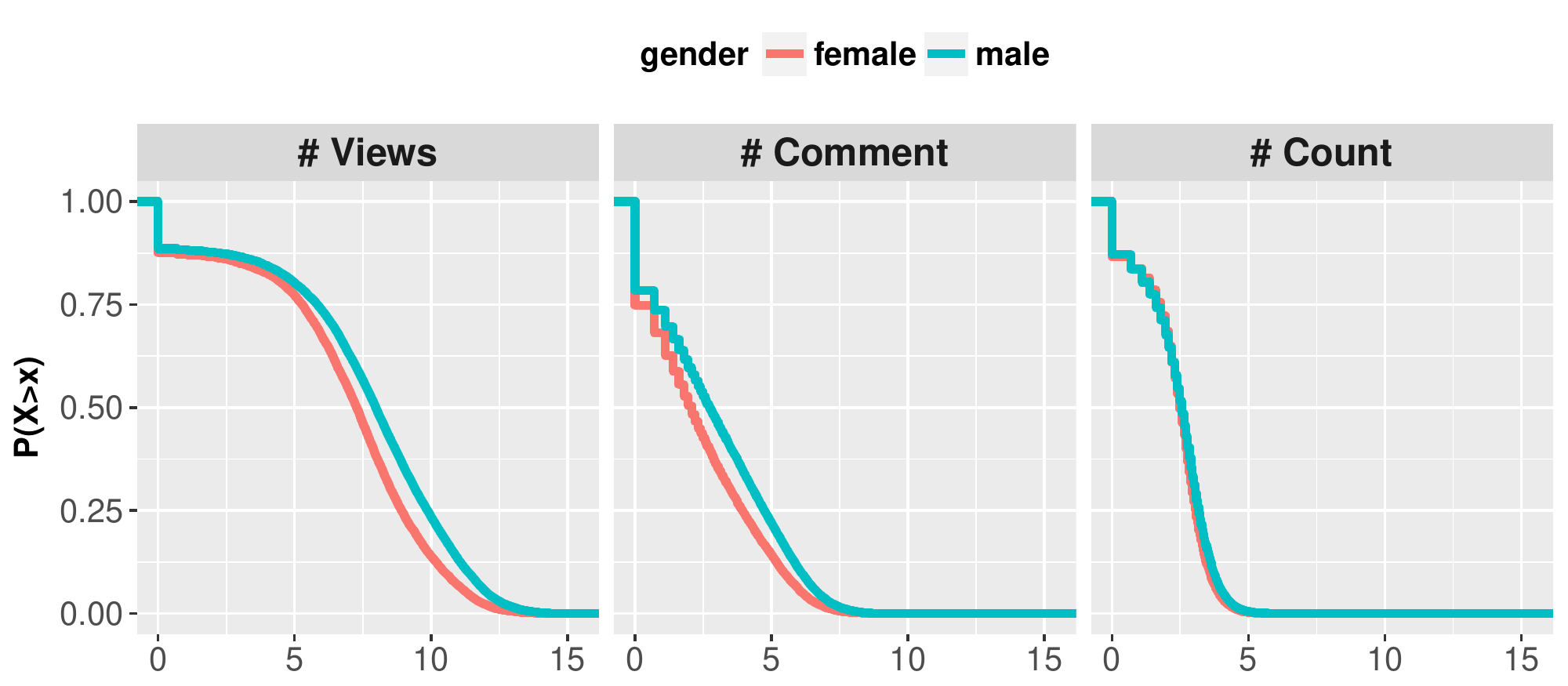}
  \vspace*{-5mm}
  \caption{Complementary cumulative distributions of project view, comment and count variables. The x-axis is in log scale.}
  \label{fig:ccdf1}
\end{figure}

Table~\ref{tbl:nb1} summarizes the predictors and overall fit of the negative binomial model using the number of appreciations as the dependent variable. The list of predictors is ordered based on their relative importance which we derived by standardizing the beta coefficients. While all predictors including top 15 topics and 10 countries were significant, we did not include them all due to their low importance and space constraint. 

Our model provides considerable explanatory power, with an improvement in deviance of 84.5K-47.3K=37.2K. Deviance is related to a model's log-likelihood and is analogous to the $R^2$ statistic for linear models. The null deviance is computed from an intercept-only model while the residual deviance is from the full model. The difference in the deviances follows a chi-squared distribution under the null hypothesis that the intercept-only model is correctly specified. $\chi^{2}$ test using the residual deviance shows that our model is significantly better in explaining the data, $\chi^{2}$(df=31\footnote{The number of degrees of freedom is given by the difference in the number of parameters in the two models, which is equal to the number of predictor variables}, N=37,777)=37.2K, $p$\textless 2e$^{-16}$. 

The number of project comments, views, and counts is the most important indicator for predicting project appreciations. Specializing in certain topics including illustration, art direction, and typography also seems to significantly impact attraction. While the number of followers is a strong measure of a user's influence, it is not the most important variable for predicting the number of appreciations. Country and gender variables are not as important as others, and thus not included in the table. However, we observed that being male suggests more appreciations, $\beta$=2.49e$^{-1}$, $p$\textless 2e$^{-16}$. To take a deeper look at the difference in gender, we calculated the complementary cumulative distribution function (CCDF) of each project-related variable by gender (Figure~\ref{fig:ccdf1}). While the number of project counts is comparable across gender, male users tend to receive more views and comments. All predictors have positive effects on increasing attraction except a few country variables; for example, users from United States and Brazil mean less appreciations. 

\subsection{R1-What attracts followers?}
Table~\ref{tbl:nb2} presents the results of our negative binomial regression model using follower count as the dependent variable; it only lists top predictors based on their relative importance. The model provides a considerable improvement in deviance, $\chi^{2}$(31, N=37.7K)=92.0K-47.5K=44.5K, $p$\textless 2e$^{-16}$. 

\begin{figure}[b]
  \includegraphics[width=\linewidth]{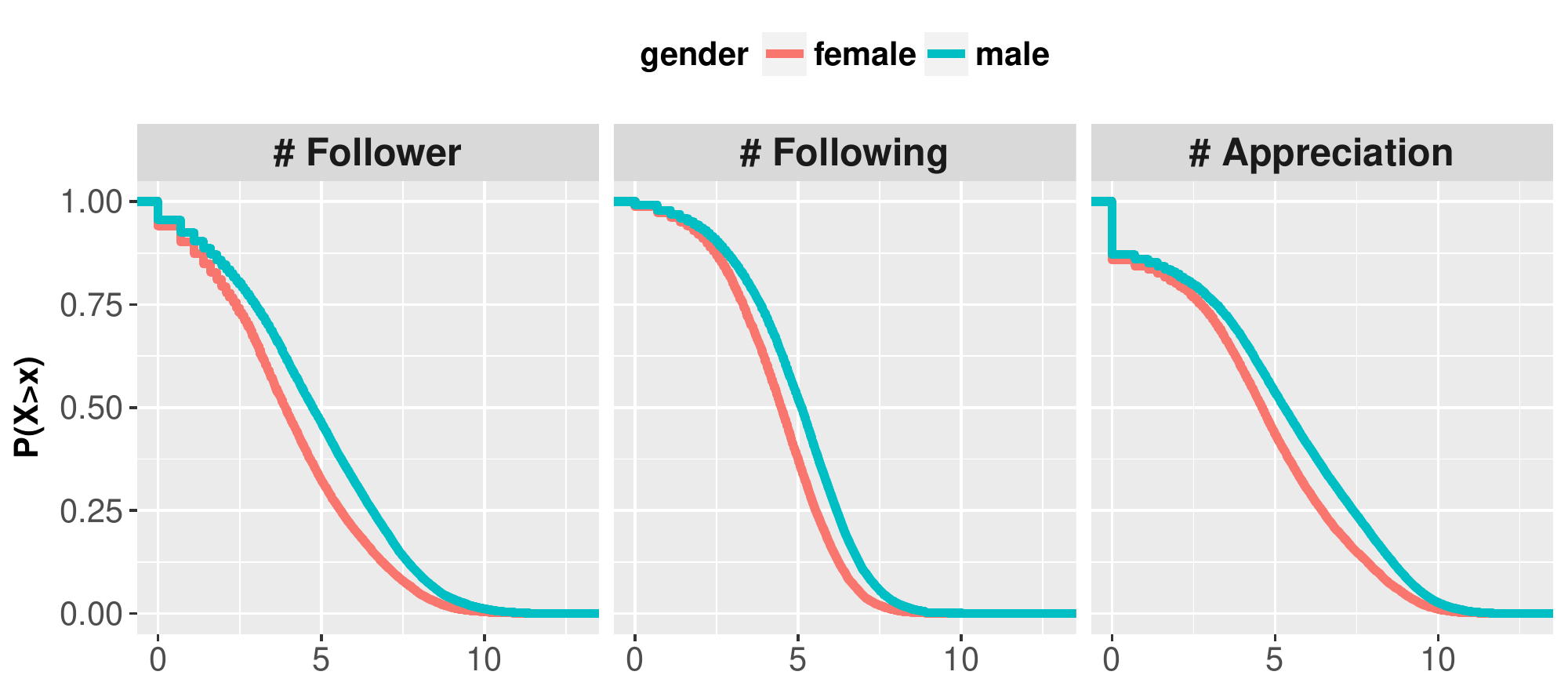}
  \vspace*{-5mm}
  \caption{Complementary cumulative distributions of follower, following, and appreciation variables. The x-axis is in log scale.  }
  \label{fig:ccdf2}
\end{figure}

Having more project appreciations leads to more followers. Residing in the United States also suggests more followers, while from India, Brazil, and Egypt indicates fewer followers,$\forall\beta$\textless 0, $\forall p$\textless 2e$^{-16}$. We also observed specializing in graphic design, which occupies the largest proportion of users, results in fewer followers ($p$\textless 2e$^{-16}$); note that users can specialize in more than one topic. Other top topics, not included in the table, that have positive effects include typography, fashion, illustration, and digital arts. Male users attract more followers. The differences in gender can be also observed from activity patterns shown in Figure~\ref{fig:ccdf1} and~\ref{fig:ccdf2}; male users follow more users and draw more attention, while both male and female users produce a similar number of projects.

\begin{table}[tb]
\centering
\caption{The result of a negative binomial regression with the number of \textit{followers} as the dependent variable. The table only lists a subset of predictors in the order of relative importance based on standardized coefficients. However, the $\beta$ coefficients here remain on their original scales.}
\label{tbl:nb2}
\begin{tabular}{@{}lllll@{}}
\toprule
Predictor & $\beta$ & std.err & z & p \\ \midrule
intercept & 4.22 & $2.17$e$^{-2}$ & 194.7 & \textless $2$e$^{-16}$ \\
appreciations & $1.47$e$^{-4}$ & $2.32$e$^{-6}$ & 63.33 & \textless $2$e$^{-16}$ \\
comments & $1.35$e$^{-3}$ & $2.37$e$^{-5}$ & 56.95 & \textless $2$e$^{-16}$ \\
following & $2.51$e$^{-4} $& $3.72$e$^{-6}$ & 67.35 & \textless $2$e$^{-16}$ \\
views & $4.20$e$^{-6}$ & $1.42$e$^{-7}$ & 29.46 & \textless $2$e$^{-16}$ \\
counts & $1.24$e$^{-2}$ & $2.20$e$^{-4}$ & 56.55 & \textless $2$e$^{-16}$ \\
gendermale & $3.32$e$^{-1}$ & $1.75$e$^{-2}$ & 18.95 & \textless $2$e$^{-16}$ \\
from united states & $4.22$e$^{-1}$ & $2.35$e$^{-2}$ & 17.98 & \textless $2$e$^{-16}$ \\
from art direction & $3.24$e$^{-1}$ & $2.11$e$^{-2}$ & 15.37 & \textless $2$e$^{-16}$ \\
from india & $-6.75$e$^{-1}$ & $4.32$e$^{-2}$ & -15.61 & \textless $2$e$^{-16}$ \\ \midrule
Summary &  null.dev &  res.dev & $\chi^{2}$  & p \\ \midrule
 & 92.0K & 47.5K & 44.5K & \textless $2$e$^{-16}$\\ \bottomrule
\end{tabular}
\end{table}

\begin{figure*}[t]
  \includegraphics[width=\textwidth]{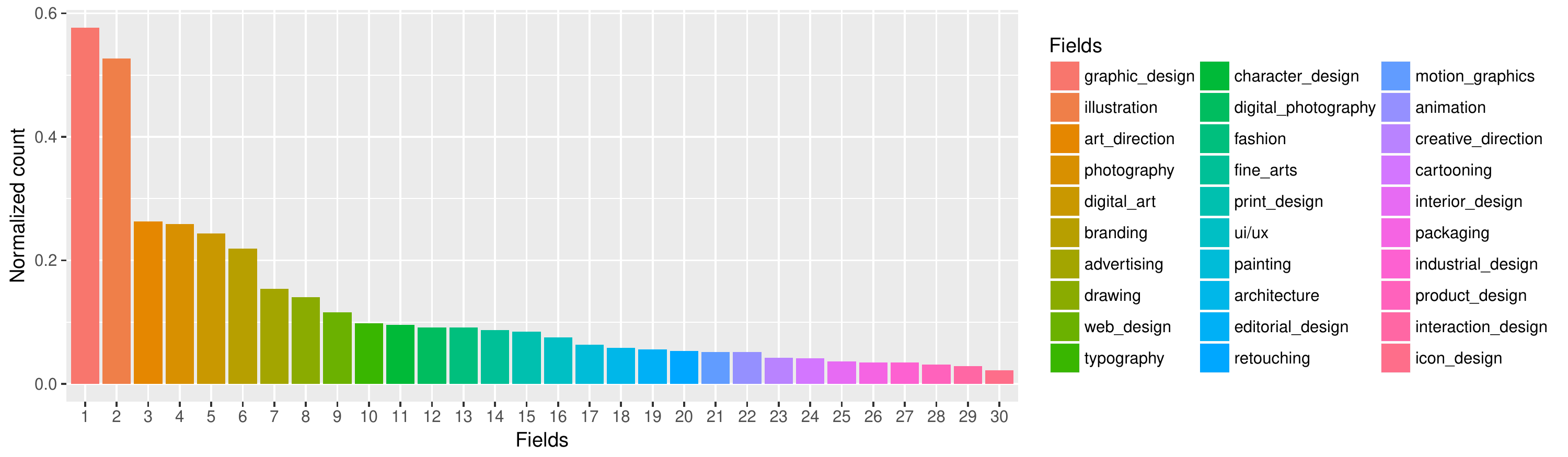}
  \vspace*{-5mm}
  \caption{Distribution of designs created by users across Behance topics. The x-axis shows a selected set of top 30 topics and the y-axis shows the proportion of designs in each category.}
  \label{fig:topics}
\end{figure*}

\subsection{R2-What topics are popular?}
We summed and normalized the topic vectors of the 668,581 projects to derive the topic vector $\hat{t}_{\boldsymbol{C}}$, where each element in the vector represents the proportion of each topic in our dataset. About 1.2\% of projects have topics that do not belong to the official 68 topics and only 0.03\% of projects did not specify topics.

Figure~\ref{fig:topics} show the overall rankings of top 30 topics, which account for more than 90\% of projects. The topic popularity follows a power law distribution ($p$\textless 0.05, using the Kolmogorov-Smirnov test). Graphic design, illustration, art direction together already cover about 33\% of projects, while industrial design, product design, interaction design, and icon design are not popular together account for less than 3\%.

For validation, we compared the topic rankings directly derived from projects with another ranking derived from users' specialization topics which are automatically computed by Behance. Using Kendall's rank correlation tau-b, the agreement between the two ranking vectors is significant ($\tau$=0.88, $p<$0.001). 

We also conducted pairwise comparisons of the topic vector $\hat{t}_{\boldsymbol{C}}$ and two additional topic vectors derived from projects created by female and male users respectively. The three rankings show a strong agreement each other ($\forall p$\textless 0.001) with some differences; the agreement between female and male users was the lowest with $\tau$=0.81. Topics in top 10 male topics but not found in top 10 female topics include fashion (M: 3.07\%, F:1.90\%) and fine arts (M: 2.79\%, F:1.84\%), while we observed web design (M: 2.17\%, F:3.02\%), typography (M: 2.37\%, F:2.37\%) in the opposite direction. For men, illustration (M: 15.2\%, F:11.78\%) is the most popular topic while graphic design (M: 14.46\%, F:13.71\%) is at the top for women.

\subsection{R2-How topics are related?}
\begin{figure}[tb]
  \includegraphics[width=\linewidth]{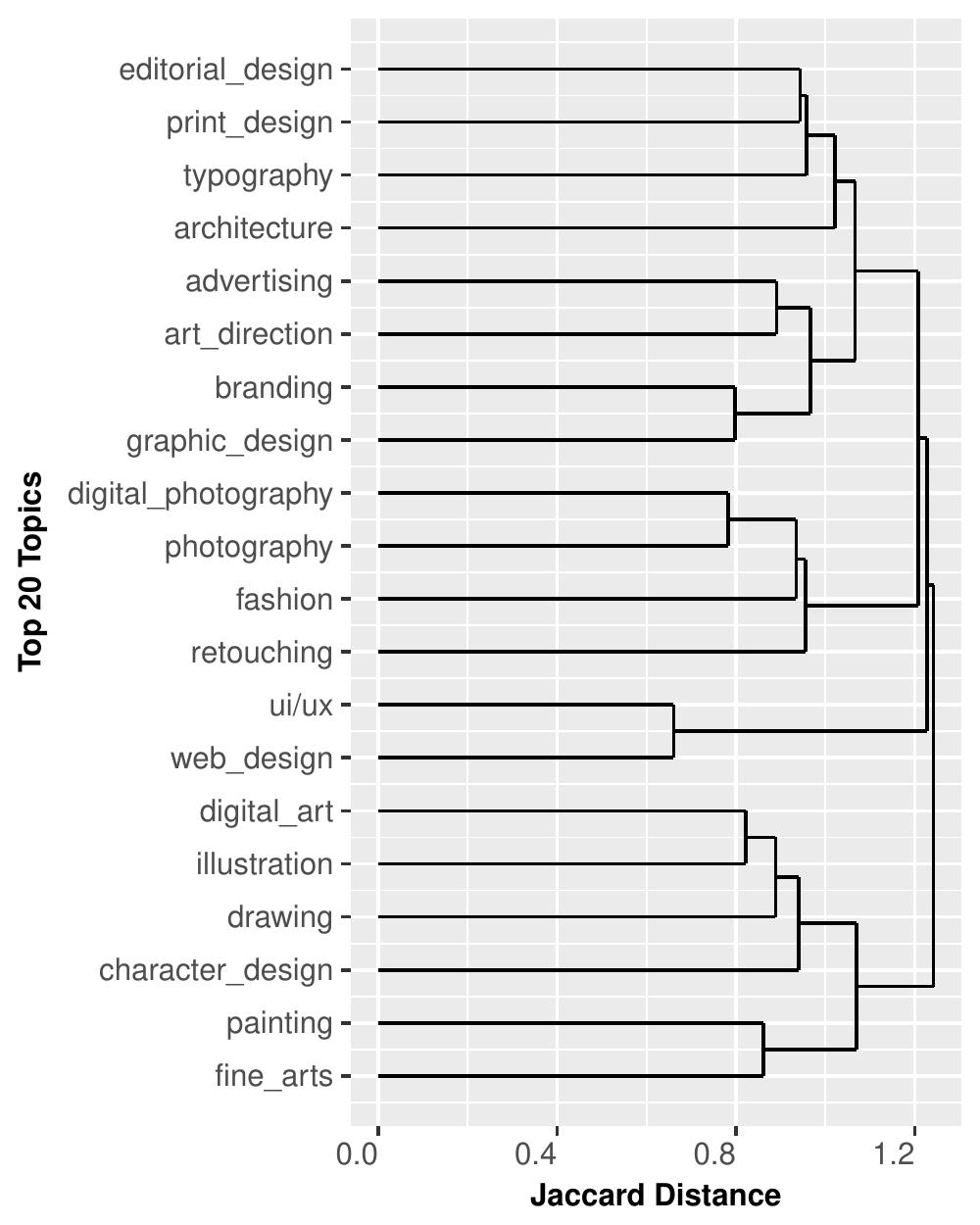}
  \vspace*{-5mm}
  \caption{Hierarchical clustering of top 20 topics}
  \label{fig:topic-cluster}
\end{figure}

Figure~\ref{fig:topic-cluster} shows the result of the hierarchical clustering of top 20 topics using Jaccard distance and Ward’s minimum criterion. The 20 topics account for more than \%80 percent of all projects. We excluded other topics as each of them only accounts for a negligible percent of projects. 

The inter-topic relationships mostly confirm our intuitions about how topics are related. The most closely related topics are web design \& UI/UX. Other closely related topics include digital photography \& photography, branding \& graphic design, digital art \& illustration, advertising \& art direction, painting \& fine arts, and editorial design \& print design in the order of similarity. 

There are 6 clusters around the Jaccard distance of 1.0, each consisting of reasonably related topics. For instance, digital photography, photography, fashion, \& retouching group together as do digital art, illustration, drawing, \& character design. Although graphic design and illustration are the two most popular topics, they do not seem to cluster together.

\subsection{R2-Are users specialists or generalists?}
By computing the entropy of each user's topic vector, we investigated the diversity of the user's interests; that is, we measured whether the user's projects are evenly distributed across different topics. The maximum entropy for a user is $ln(68)$=4.22 when the user created an equal number of projects in each topic.

 Similar to \citeauthor{chang2014specialization}, we divided users into three equal-sized groups based on project counts and compared the average topic entropy across the groups with the hypothesis that a user with a larger number of projects has a diverse interest. 
 
\begin{figure}[tb]
  \includegraphics[width=\linewidth]{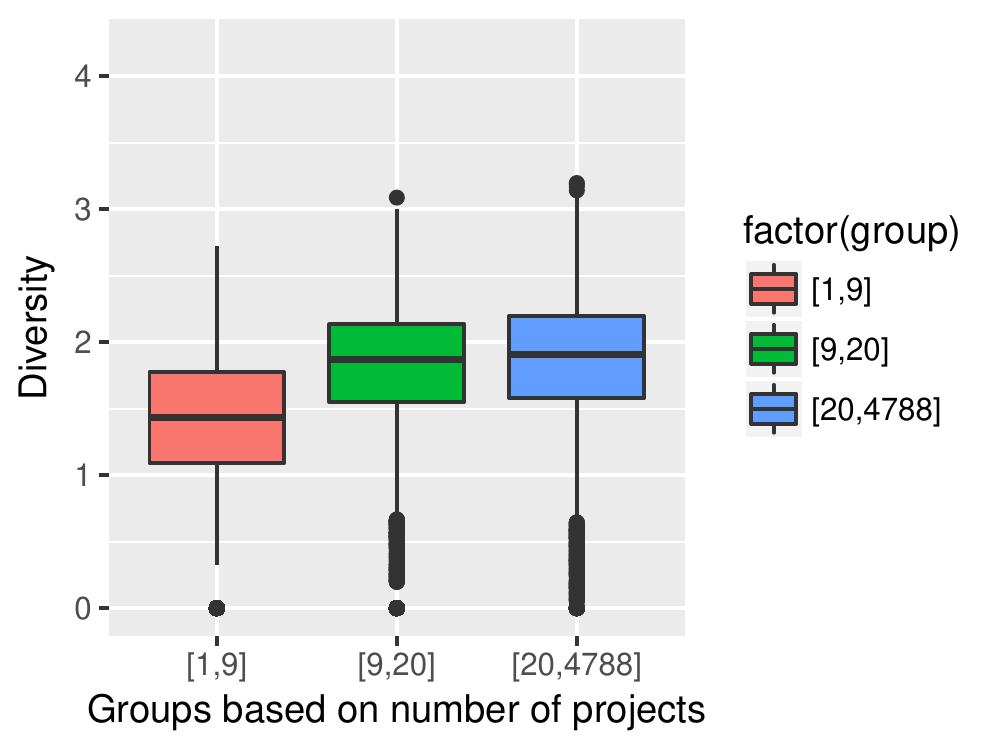}
  \vspace*{-5mm}
  \caption{Topic diversity (entropy) of three equal-sized groups of users based on project counts.}
  \label{fig:diversity}
\end{figure}

 Figure~\ref{fig:diversity} shows the entropy distributions for the three groups. The differences between the groups were significant, $\forall p$\textless 0.001. The effect size between the first and second groups is 0.84 (large) while the effect size between the second and third groups is 0.11 (negligible). We did not find any significant difference across gender in general.
 
 Overall, the mean entropy values of the groups all are less than 1.95. This value is the maximum entropy when there are only 7 topics. We also observed the mean entropy of all users is 1.67. Given the fact that the maximum entropy of 68 topics is 4.22, the overall topic diversity seems relatively low, suggesting that most users are specialists while some differences may exist. 
 
 For example, in the third group, we found a generalist (entropy=2.60, project count: 619) who created projects in 48 topics. However, 71.5\% of the projects are covered by only 6 topics which include graphic design, branding, web design, UI/UX, calligraphy, and typography; in addition, some of the 6 topics are related to each other such as web design and UI/UX.
 
 \ignore{
 We also found a significant difference across gender ($p<$0.05), suggesting that female users work on more diverse topics. However, Cohen's effect size ($d$=.038) value suggests a negligible practical significance. Among the three groups, there is only a significant difference between the second and third groups ($p<$0.05), but the effect size is also very small ($d$=.047).
}
 
\subsection{R2-Is there homophily in appreciating others?}
The average homophily of all users is 0.91 (Std=0.08) where we computed each homophily of a user through the cosine similarity between the two sets of topics used in the projects created ($\hat{u}_{\boldsymbol{P_u}}$) and appreciated ($\hat{u}_{\boldsymbol{A_u}}$) by the user. 

\begin{figure}[tb]
  \includegraphics[width=\linewidth]{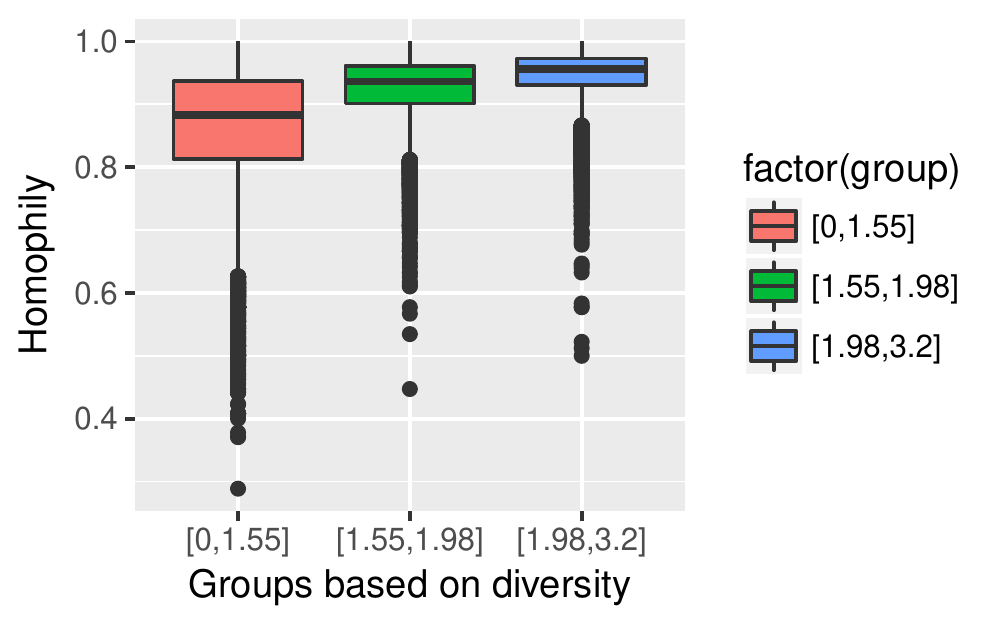}
  \vspace*{-5mm}
  \caption{Homophily (cosine similarity of $\hat{u}_{\boldsymbol{P_u}}$ and $\hat{u}_{\boldsymbol{A_u}}$) of three equal-sized groups of users based on topic diversity.}
  \label{fig:homophily}
\end{figure}

We initially hypothesized that users with high topic diversity would have higher tendency to appreciate projects of diverse topics, thus resulting in lower homophily. We divided users into three equal-sized groups based on topic diversity and plotted the homophily distribution per each group (Figure~\ref{fig:homophily}). The graph shows an opposing trend of our hypothesis (differences among groups are significant ($\forall p<$0.001, $d_{\text{group1},\text{group2}}$=0.78, $d_{\text{group2},\text{group3}}$=0.45). We observed that the same trend holds when we divided users into five groups ($p<$0.001). 

We also tested whether men and women differed in the homophily of their interests. Overall, there is a significant difference across gender ($p<$0.001), suggesting that women tend to create and appreciate projects of similar topics to a greater extent than men. However, Cohen's effect size ($d$=0.067) value suggests a negligible practical significance. Among the three groups, there is only a significant difference between the first and second groups ($p<$0.001), but the effect size is also very small ($d$=0.10).

\subsection{R3-What colors attract appreciations?}

\begin{table}[tb]
\centering
\caption{The result of a negative binomial regression model for color analysis. The table only lists a subset of predictors in the order of relative importance based on standardized coefficients. The $\beta$ coefficients here remain on their original scales.}
\label{tbl:nb-color}
\begin{tabular}{@{}lllll@{}}
\toprule
Predictor & Estimate & std.err & z & p \\ \midrule
intercept & 4.02 & $1.33$e$^{-2}$ & 302.9 & \textless $2$e$^{-16}$ \\
followers & $1.83$e$^{-4}$ & $2.16$e$^{-7}$ & 848.9 & \textless $2$e$^{-16}$ \\
white & $-5.98$e$^{-1}$ & $1.86$e$^{-2}$ & -32.09 & \textless $2$e$^{-16}$ \\
gray & $-4.18$e$^{-1}$ & $1.57$e$^{-2}$ & -26.63 & \textless $2$e$^{-16}$ \\
value & $1.28$e$^{-3}$ & $9.11$e$^{-5}$ & 14.10 & \textless $2$e$^{-16}$ \\
saturation & $-1.08$e$^{-3}$ & $7.83$e$^{-5}$ & -13.79 & \textless $2$e$^{-16}$ \\
colorfulness1 & $-1.34$e$^{-3}$ & $1.21$e$^{-4}$ & -11.05 & \textless $2$e$^{-16}$ \\
black & $-1.75$e$^{-1}$ & $1.95$e$^{-2}$ & -9.00 & \textless $2$e$^{-16}$ \\
olive & $-5.97$e$^{-1}$ & $4.88$e$^{-2}$ & -12.23 & \textless $2$e$^{-16}$ \\
colorfulness2 & $1.07$e$^{-3}$ & $1.09$e$^{-4}$ & 9.81 & \textless $2$e$^{-16}$ \\ \midrule
Summary & null.dev & res.dev & $\chi^{2}$ & p \\\midrule
 & 859K & 680K & 179K & \textless $2$e$^{-16}$ \\ \bottomrule
\end{tabular}
\end{table}

Table~\ref{tbl:nb-color} presents the result of a negative binomial regression model using the color metrics as experimental variables. The reduction in deviance from the full model to the null model is significant, $\chi^{2}$(21, N=542K)=859K-680K=179K, $p$\textless 2e$^{-16}$. 

We also compared the full model with a simpler model using the control variable alone (number of followers). The AIC\footnote{AIC=2$k$-2$\ln({\hat{L}})$, where $\hat{L}$ is the maximum value of the likelihood function of the model and $k$ is the number of estimated parameters.} value of the full model was lower than that of the control-only model (1.90K$>$0), meaning that the full model has a better fit. The chi-square test comparing the two models indicates that the color variables are significant predictors, $\chi^{2}$(20), N=542K)=1.90K, $p$\textless 2e$^{-16}$.

 Among all color variables, the highest effect on appreciation is found in white color, $\beta$=-5.98e$^{-1}$, $p$\textless 2e$^{-16}$. Gray color, saturation, and colorfulness1~\cite{hasler2003measuring} had a negative effect while colorfulness2~\cite{yendrikhovskij1998optimizing} and value (brightness) has a positive effect on appreciation.

Other significant predictors not included in the table are red, aqua, teal, and lime colors in the order of their standardized contribution. The $\beta$ coefficients for red and lime colors are negative, while aqua and teal increases the chance of receiving appreciations. Silver, maroon, purple, fuchsia, green, yellow, navy, and blue do not seem to impact appreciation.

\section{Discussion}
In this section, we discuss our findings in light of research questions. We also qualitatively compare our results with previous research on Pinterest. 

\subsection{R1-Activity: What attract attention?}
Our regression models capture 44\% of the variance around the number of appreciations and 48\% of the variance around the number of followers. The models provide a fairly good explanatory power although additional features would further improve prediction accuracy. 

With regard to predicting appreciations, the number of project comments, views, and counts is a strong indicator. While comments and views are out of a user's control, this indicates that the user may produce more works to be more recognized. Most predictors also showed expected behavior; we see specializing in popular topics, followers, and followees all contribute. Interestingly, we see that being from the U.S has a negative impact on appreciation, $\beta$=$-1.03$e$^{-1}$, $p$\textless 2e$^{-16}$.

For predicting followers, the number of appreciations, comments, views, counts also highly contributes so as the number of followees. Focusing on popular topics is also helpful except graphic design which we found has a negative impact, $\beta$=$-1.85$e$^{-1}$, $p$\textless 2e$^{-16}$. Users from the United States are likely to attract more followers; this result conforms to the follower model on Pinterest by~\citeauthor{gilbert2013need}.

What is more surprising is that male users attract more appreciations and followers than female users. Similarly, \citeauthor{gilbert2013need} found that being female users suggest fewer followers on Pinterest. Our result might be due to the larger population of male users on Behance. Further studies are necessary for a deeper understanding of the effect of gender. 
 
\subsection{R2-Topic: How people engage with various topics?}
Our topic rankings are community-based, derived by aggregating user-specified topics for projects. The two most popular topics are graphic design and illustration. Since they are generally used for any visual communication, we expect that they are specified along with many other topics. 

We see some discrepancy between top 12 topics in our rankings and 12 official popular topics from Behance which had 6 different topics, including interaction design, industrial design, motion graphics, fashion, architecture, and UI/UX, while we had digital art, advertising, drawing, typography, character design, and digital photography. While this discrepancy might be due to the lack of samples we have in our data, Behance seems to select topics that are popular and distinctive at the same time. 

The overall entropy of all users (M=1.67) is lower than the mean entropy (M~2.02, estimated) that ~\citeauthor{chang2014specialization} found on Pinterest data which even has a lower number of categories (33, $ln(32)$=3.50 vs 68 topics, $ln(68)$=4.22). This suggests that Behance users are more specialized. We did not find any significant difference across gender, which also contrasts with the result from~\citeauthor{chang2014specialization} where they found that women curate significantly more diverse content than men. 

A high similarity between two sets of topics, one from projects created by users and another from those appreciated by the users, suggests that there is homophily among users. While we did not directly analyze inter-user relationships (e.g., topics from followers and followees), this is a reasonable proxy for homophily as the projects appreciated are created by different users. We believe homophily of interests among creative professionals is an interesting research direction and needs further investigations; for example, do appreciations from people who specialize in different topics carry a different degree of recognition? what about appreciations from different gender?

\subsection{R3-Color: What colors attract appreciations?}
Our analysis on colors shows that grayscale colors (black, white, and gray) have a negative impact on attracting appreciation. This corresponds to the result of similar color analysis on Pinterest~\cite{bakhshi2015red} where black \& white images are not shared as frequently as colored images. Similarly, the effect of lightness (value in the HSV color space) indicates that the more brighter an image is, the more appreciation it can expect, $\beta$=$1.28$e$^{-3}$, $p<$2e$^{-16}$. This also conforms to previous research showing that people consider bright objects as good, whereas dark objects are bad~\cite{meier2004good}.

On the other hand, we observed mixed effects of colors that have non-saturated hues. Only aqua and teal, which lie at 180\textdegree in the hue dimension, shows a positive impact on appreciation. Red, olive, lime, and green colors, which covers 0\textdegree to 120\textdegree have the opposite effect, while other colors have no effect. Our result is not consistent with the result on Pinterest~\cite{bakhshi2015red} where red, purple, and pink colors promotes diffusion of content (i.e, repins). Further studies will be necessary to understand the differences. The effect of other color features would be also interesting to investigate~\cite{khosla2014makes}.

\section{Limitations}
Our work has a number of limitations. It is possible that our sample may not represent the actual population on Behance; we sampled data twice and briefly, confirmed that our major findings hold. In addition, our analysis depends on a commercial service for predicting gender. While we manually verified a few sample of predictions, we did not confirm its true accuracy. Because of the gender prediction, we happened to exclude some Asian countries to which our results may not generalize. For color analysis, we used web safe colors that might not appropriate for the images of artworks.

\section{Conclusion}
In this paper, we provided a statistical overview of Behance, a social network site for creative professionals. We found several significant results. First, being male leads to more followers and appreciations. Second, most users are specialists focusing on a few topics and they tend to appreciate projects in their specialized topics. Finally, grayscale colors suggest less appreciations. For future work, we plan to investigate another creative community site, Dribble, to see if our findings hold on the site as well. 

\section{Acknowledgement}
This work has also been made possible through support the Kwanjeong Educational Foundation.
\newpage
\ignore{
The users are mostly individuals while some are organizations or teams. Belong to a group will help users become more popular attracting more attention?

Design implications for online community for creative professionals. How to design features that promote collaboration and interaction across different topics? How the community can help improving the creativity?
collection vs projects

Not all activites are happening on Behance but distributed to other third-party online galleries and social networks.
}

\bibliography{paper}

\begin{thebibliography}{}

\bibitem[\protect\citeauthoryear{Akdag~Salah and Salah}{2013}]{AkdagSalah2013}
Akdag~Salah, A.~A., and Salah, A.~A.
\newblock 2013.
\newblock Flow of innovation in deviantart: following artists on an online
  social network site.
\newblock {\em Mind {\&} Society} 12(1):137--149.

\bibitem[\protect\citeauthoryear{Bakhshi and Gilbert}{2015}]{bakhshi2015red}
Bakhshi, S., and Gilbert, E.
\newblock 2015.
\newblock Red, purple and pink: The colors of diffusion on pinterest.
\newblock {\em PloS one} 10(2):e0117148.

\bibitem[\protect\citeauthoryear{Cameron and
  Trivedi}{2013}]{cameron2013regression}
Cameron, A.~C., and Trivedi, P.~K.
\newblock 2013.
\newblock {\em Regression analysis of count data}, volume~53.
\newblock Cambridge university press.

\bibitem[\protect\citeauthoryear{Chang \bgroup et al\mbox.\egroup
  }{2014a}]{chang2014specialization}
Chang, S.; Kumar, V.; Gilbert, E.; and Terveen, L.~G.
\newblock 2014a.
\newblock Specialization, homophily, and gender in a social curation site:
  findings from pinterest.
\newblock In {\em Proceedings of the 17th ACM conference on Computer supported
  cooperative work \& social computing},  674--686.
\newblock ACM.

\bibitem[\protect\citeauthoryear{Chang \bgroup et al\mbox.\egroup
  }{2014b}]{chang2014tumblr}
Chang, Y.; Tang, L.; Inagaki, Y.; and Liu, Y.
\newblock 2014b.
\newblock What is tumblr: A statistical overview and comparison.
\newblock {\em ACM SIGKDD Explorations Newsletter} 16(1):21--29.

\bibitem[\protect\citeauthoryear{Deka \bgroup et al\mbox.\egroup
  }{2015}]{deka2015ranking}
Deka, B.; Yu, H.; Ho, D.; Huang, Z.; Talton, J.~O.; and Kumar, R.
\newblock 2015.
\newblock Ranking designs and users in online social networks.
\newblock In {\em Proceedings of the 33rd Annual ACM Conference Extended
  Abstracts on Human Factors in Computing Systems},  1887--1892.
\newblock ACM.

\bibitem[\protect\citeauthoryear{Florida}{2012}]{florida2012rise}
Florida, R.
\newblock 2012.
\newblock The rise of the creative class: Revisited.
\newblock {\em New York}.

\bibitem[\protect\citeauthoryear{Gilbert \bgroup et al\mbox.\egroup
  }{2013}]{gilbert2013need}
Gilbert, E.; Bakhshi, S.; Chang, S.; and Terveen, L.
\newblock 2013.
\newblock I need to try this?: a statistical overview of pinterest.
\newblock In {\em Proceedings of the SIGCHI conference on human factors in
  computing systems},  2427--2436.
\newblock ACM.

\bibitem[\protect\citeauthoryear{Halstead, Serrano, and
  Proctor}{2015}]{halstead2015finding}
Halstead, S.; Serrano, H.~D.; and Proctor, S.
\newblock 2015.
\newblock Finding top ui/ux design talent on adobe behance.
\newblock {\em Procedia Computer Science} 51:2426--2434.

\bibitem[\protect\citeauthoryear{Han \bgroup et al\mbox.\egroup
  }{2014}]{han2014collecting}
Han, J.; Choi, D.; Chun, B.-G.; Kwon, T.; Kim, H.-c.; and Choi, Y.
\newblock 2014.
\newblock Collecting, organizing, and sharing pins in pinterest:
  Interest-driven or social-driven?
\newblock {\em SIGMETRICS Perform. Eval. Rev.} 42(1):15--27.

\bibitem[\protect\citeauthoryear{Hasler and
  Suesstrunk}{2003}]{hasler2003measuring}
Hasler, D., and Suesstrunk, S.~E.
\newblock 2003.
\newblock Measuring colorfulness in natural images.
\newblock {\em Proc. SPIE} 5007:87--95.

\bibitem[\protect\citeauthoryear{Hu \bgroup et al\mbox.\egroup
  }{2014}]{hu2014we}
Hu, Y.; Manikonda, L.; Kambhampati, S.; et~al.
\newblock 2014.
\newblock What we instagram: A first analysis of instagram photo content and
  user types.
\newblock In {\em ICWSM}.

\bibitem[\protect\citeauthoryear{Jones}{2015}]{jones2015collective}
Jones, B.~L.
\newblock 2015.
\newblock Collective learning resources: Connecting social-learning practices
  in deviantart to art education.
\newblock {\em Studies in Art Education} 56(4):341--354.

\bibitem[\protect\citeauthoryear{Khosla, Das~Sarma, and
  Hamid}{2014}]{khosla2014makes}
Khosla, A.; Das~Sarma, A.; and Hamid, R.
\newblock 2014.
\newblock What makes an image popular?
\newblock In {\em Proceedings of the 23rd International Conference on World
  Wide Web}, WWW '14,  867--876.
\newblock New York, NY, USA: ACM.

\bibitem[\protect\citeauthoryear{Kwak \bgroup et al\mbox.\egroup
  }{2010}]{kwak2010twitter}
Kwak, H.; Lee, C.; Park, H.; and Moon, S.
\newblock 2010.
\newblock What is twitter, a social network or a news media?
\newblock In {\em Proceedings of the 19th International Conference on World
  Wide Web}, WWW '10,  591--600.
\newblock New York, NY, USA: ACM.

\bibitem[\protect\citeauthoryear{Leskovec and
  Faloutsos}{2006}]{leskovec2006sampling}
Leskovec, J., and Faloutsos, C.
\newblock 2006.
\newblock Sampling from large graphs.
\newblock In {\em Proceedings of the 12th ACM SIGKDD international conference
  on Knowledge discovery and data mining},  631--636.
\newblock ACM.

\bibitem[\protect\citeauthoryear{Meier, Robinson, and
  Clore}{2004}]{meier2004good}
Meier, B.~P.; Robinson, M.~D.; and Clore, G.~L.
\newblock 2004.
\newblock Why good guys wear white automatic inferences about stimulus valence
  based on brightness.
\newblock {\em Psychological science} 15(2):82--87.

\bibitem[\protect\citeauthoryear{Ottoni \bgroup et al\mbox.\egroup
  }{2013}]{ottoni2013ladies}
Ottoni, R.; Pesce, J.~P.; Casas, D.~L.; Jr., G.~F.; Jr., W.~M.; Kumaraguru, P.;
  and Almeida, V.
\newblock 2013.
\newblock Ladies first: Analyzing gender roles and behaviors in pinterest.
\newblock In {\em International AAAI Conference on Web and Social Media}.

\bibitem[\protect\citeauthoryear{Perkel}{2011}]{perkel2011making}
Perkel, D.
\newblock 2011.
\newblock Making art, creating infrastructure: Deviantart and the production of
  the web.
\newblock In {\em Ph.D. dissertation}.

\bibitem[\protect\citeauthoryear{Reinecke \bgroup et al\mbox.\egroup
  }{2013}]{reinecke2013predicting}
Reinecke, K.; Yeh, T.; Miratrix, L.; Mardiko, R.; Zhao, Y.; Liu, J.; and Gajos,
  K.~Z.
\newblock 2013.
\newblock Predicting users' first impressions of website aesthetics with a
  quantification of perceived visual complexity and colorfulness.
\newblock In {\em Proceedings of the SIGCHI Conference on Human Factors in
  Computing Systems}, CHI '13,  2049--2058.
\newblock New York, NY, USA: ACM.

\bibitem[\protect\citeauthoryear{Rudolph, Hoffman, and
  Hertzmann}{2016}]{rudolph2016joint}
Rudolph, M.~R.; Hoffman, M.; and Hertzmann, A.
\newblock 2016.
\newblock A joint model for who-to-follow and what-to-view recommendations on
  behance.
\newblock In {\em Proceedings of the 25th International Conference Companion on
  World Wide Web},  581--584.
\newblock International World Wide Web Conferences Steering Committee.

\bibitem[\protect\citeauthoryear{Salah \bgroup et al\mbox.\egroup
  }{2012}]{salah2012deviantart}
Salah, A.~A.; Salah, A.~A.; Buter, B.; Dijkshoorn, N.; Modolo, D.; Nguyen, Q.;
  van Noort, S.; and van~de Poel, B.
\newblock 2012.
\newblock Deviantart in spotlight: A network of artists.
\newblock {\em Leonardo} 45(5):486--487.

\bibitem[\protect\citeauthoryear{Salah}{2010}]{salah2010online}
Salah, A.~A.
\newblock 2010.
\newblock The online potential of art creation and dissemination: deviantart as
  the next art venue.
\newblock In {\em Proceedings of the 2010 international conference on
  Electronic Visualisation and the Arts},  16--22.
\newblock British Computer Society.

\bibitem[\protect\citeauthoryear{Scolere and
  Humphreys}{2016}]{scolere2016pinning}
Scolere, L., and Humphreys, L.
\newblock 2016.
\newblock Pinning design: The curatorial labor of creative professionals.
\newblock {\em Social Media + Society} 2(1):2056305116633481.

\bibitem[\protect\citeauthoryear{Tan and Yuen}{2015}]{tan2015destuckification}
Tan, Y.~Y., and Yuen, A.~H.
\newblock 2015.
\newblock “destuckification”: Use of social media for enhancing design
  practices.
\newblock In {\em New Media, Knowledge Practices and Multiliteracies}.
  Springer.
\newblock  67--75.

\bibitem[\protect\citeauthoryear{Yendrikhovskij, Blommaert, and de
  Ridder}{1998}]{yendrikhovskij1998optimizing}
Yendrikhovskij, S.; Blommaert, F.~J.; and de~Ridder, H.
\newblock 1998.
\newblock Optimizing color reproduction of natural images.
\newblock In {\em Color and Imaging Conference},  140--145.
\newblock Society for Imaging Science and Technology.

\end{thebibliography}
\bibliographystyle{aaai}
\end{document}